\documentclass[preprint,superscriptaddress,tightenlines,nofootinbib]{revtex4}

\pdfoutput=1
\usepackage{graphicx}% Include figure files
\usepackage{dcolumn}% Align table columns on decimal point
\usepackage{bm}% bold math
\usepackage{xcolor}
\usepackage{amsmath}
\usepackage{tabularx}
\usepackage{tikz}
\usepackage{hyperref}
\usetikzlibrary{arrows}

\newcommand{\beq}{\begin{equation}}
\newcommand{\eeq}{\end{equation}}
\newcommand{\beqn}{\begin{eqnarray}}
\newcommand{\eeqn}{\end{eqnarray}}

{}

\newcommand{\Dn}{\ensuremath{D^0}}
\newcommand{\Db}{\ensuremath{\overline{D}{}^0}}

\newcommand{\ddb}{\Dn-\Db}

\newcommand{\Kn}{\ensuremath{K^0}}
\newcommand{\Kb}{\ensuremath{\overline{K}{}^0}}
\newcommand{\Kba}{\ensuremath{\left|\Kb(t)\right\rangle}}
\newcommand{\Kna}{\ensuremath{\left|\Kn(t)\right\rangle}}
\newcommand{\Ksa}{\ensuremath{\left| K_{S} \right\rangle}}
\newcommand{\Kla}{\ensuremath{\left| K_{L} \right\rangle}}

\newcommand{\Dna}{\ensuremath{\left|\Dn\right\rangle}}
\newcommand{\Dba}{\ensuremath{\left|\Db\right\rangle}}

\newcommand{\pqk}{\ensuremath{\left( \frac{p}{q} \right)_{\!\!K}}}
\newcommand{\qpd}{\ensuremath{\left( \frac{q}{p} \right)_{\!\!D}}}
\newcommand{\pqd}{\ensuremath{\left( \frac{p}{q} \right)_{\!\!D}}}

\newcommand{\Ks}{\ensuremath{K^0_S}}
\newcommand{\Kl}{\ensuremath{K^0_L}}
\newcommand{\pp}{\ensuremath{\pi^+ \pi^-}}

\newcommand{\elas}{\ensuremath{e^{-i\lambda_{S} t}}}

\newcommand{\elal}{\ensuremath{e^{-i\lambda_{L} t}}}

\newcommand{\egs}{\ensuremath{e^{-\Gamma_{S} t}}}
\newcommand{\egl}{\ensuremath{e^{-\Gamma_{L} t}}}

\newcommand{\egsum}{\ensuremath{e^{-\frac{1}{2}(\Gamma_{L}+\Gamma_{S})t}}}
\newcommand{\srd}{\ensuremath{\sqrt{r_{f}}}}
\newcommand{\srdbar}{\ensuremath{\sqrt{\overline{r_{f}}}}}
\newcommand{\phipm}{\ensuremath{\phi_{+-}}}
\newcommand{\etpm}{\ensuremath{\eta_{+-}}}
\newcommand{\sut}{\ensuremath{SU(3)_f}}

\newcommand{\ctau}{\ensuremath{c}--\ensuremath{\tau}}

\begin{document}

\title{\boldmath Time-dependent study of $K_{S} \to \pi^{+} \pi^{-}$ decays for flavour physics measurements}

\author{P. Pakhlov}
\email{pakhlov@lebedev.ru}
\address{P.N. Lebedev Physical Institute of the RAS, Moscow, Russia}
\address{Higher School of Economics (National Research University), Moscow, Russia}

\author{V. Popov }
\email{popovve@lebedev.ru}
\address{P.N. Lebedev Physical Institute of the RAS, Moscow, Russia}
%\href{mailto:popovve@lebedev.ru}

\date{\today}

\begin{abstract}
Nowadays High Energy Physics experiments can accumulate unprecedented statistics of heavy flavour decays that allows to apply new methods, based on the study of very rare phenomena, which used to be just desperate. In this paper we propose a new method to measure composition of \Kn-\Kb, produced in a decay of heavy hadrons. This composition contains important information, in particular about weak and strong phases between amplitudes of the produced \Kn\ and \Kb. We consider possibility to measure these parameters with time-dependent $\Kn \to \pp$ analysis. Due to $CP$-violation in kaon mixing time-dependent decay rates of \Kn\ and \Kb\ differ, and the initial amplitudes revealed in the $CP$-violating decay pattern. We perform phenomenological study of \Kn\ decay evolution initially produced as a combination $a \Kna + b \Kba$, where $a$ and $b$, complex amplitudes, could also be dependent on decay time of heavy mother particle. In particular we consider cases of charmed hadrons decays: $D^+ \to \Kn \pi^+$, $D_s^+ \to \Kn K^+$, $\Lambda \to p \Kn$ and with some assumptions $\Dn \to \Kn \pi^0$. This can be used to test the sum rule for charmed mesons and to obtain input for the full constraint of the two body amplitudes of $D$-mesons. 
\end{abstract}

\maketitle

\section{Introduction}

The final states of heavy hadron decays sometimes differ only by the replacement of the strange \Kn\ meson by its antiparticle due to the contribution of different diagrams or neutral meson mixing, thus representing a superposition of the strange and antistrange states. An example is the decay $D^+ \to \Kb(\Kn) \pi^+$, where both \Kb\ and \Kn\ are produced due to the presence of both Cabibbo favourite (CF) and doubly Cabibbo suppressed (DCS) amplitudes. 

\begin{figure}[bht]
    \centering
    \begin{tikzpicture}
    \draw[-triangle 45, thick] (0,0) -- (3., 4) ;
    \draw (1., 2.5) node[rotate=55] {$\mathbf{\sqrt{2} A_{D^0 \to \overline{K}^0 \pi^0}}$};
    \draw[-triangle 45, thick] (3., 4) -- (4., 0);
    \draw (4.1, 2.1) node[rotate=-73] {$\mathbf{A_{D^0 \to K^- \pi^+}}$};
    \draw[-triangle 45, thick] (0,0) -- (4., 0);
    \draw (2., -0.5) node {$\mathbf{A_{D^+ \to \overline{K}^0 \pi^+}}$};

    \draw[-triangle 45, thick] (7.5,0) -- (9, 4) ;
    \draw[-triangle 45, thick] (9, 4) -- (13, 3.5) ;
    \draw[-triangle 45, thick] (13,3.5) -- (11.5, 0.55) ;
    \draw[triangle 45-, thick] (11.5,0.55) -- (7.5, 0) ;

    \draw (7.5, 2.) node[rotate=72] {$\mathbf{\sqrt{2} A_{D^0 \to K^0 \pi^0}}$};
    \draw (10.9, 4.2) node[rotate=-8] {$\mathbf{A_{D^0 \to K^+ \pi^-}}$};
    \draw (13.3, 1.2) node[rotate=0] {$\mathbf{\sqrt{2} A_{D^+ \to K^+ \pi^0}}$};
    \draw (9.25, -0.5) node[rotate=0] {$\mathbf{A_{D^+ \to K^0 \pi^+}}$};
    
    \draw[-triangle 45, dashed] (7.5,0) -- (9.3, 2.4) ;
    \draw[-triangle 45, dashed] (9., 4) -- (10.25, 2) ;
    \draw[-triangle 45, dashed] (7.5, 0) -- (11, 0) ;
    \draw (8.124,1.7) .. controls (8.4,1.9) and (8.6,1.8) .. (8.72, 1.6);
    \draw (9.8,3.91) .. controls (9.9,3.7) and (9.9, 3.3) .. (9.5, 3.2);
    \draw (10.5, 0) .. controls (10.7, 0.175) and (10.7, 0.35) .. (10.5, 0.425);
    \draw (8.7, 2.2) node {$\delta^{00}$};
    \draw (10.4, 3.3) node {$\delta^{+-}$};
    \draw (11.6, 0.2) node {$\delta^{0+}$};
    
    \draw (0, 4.) node {$\mathrm{(a)}$};
    \draw (7.5, 4) node {$\mathrm{(b)}$};
\end{tikzpicture}
\caption{$SU_f(3)$ sum rules for CF {\it (a)} and DCS {\it (b)} decay amplitudes.}
\label{fig:SumRules}
\end{figure}
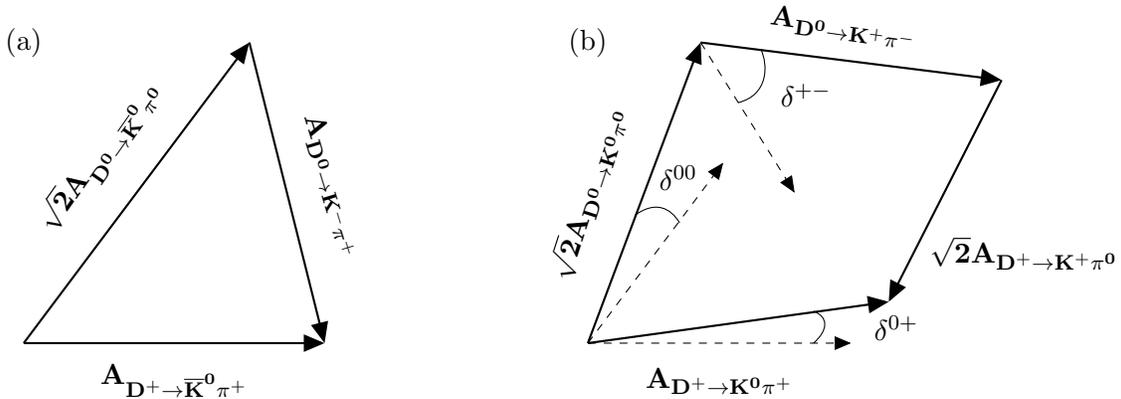

Considering charm hadron decays from the point of view of flavour \sut\  symmetry one can obtain the following sum rules~\cite{Grossman:2012ry}: 
\beqn
\sqrt{2} A_{\Dn \to \Kb \pi^0} + A_{\Dn \to K^- \pi^+} - A_{D^+ \to \Kb \pi^+} = 0, 
\label{eq:CF_sum} \\
\sqrt{2} A_{\Dn \to \Kn \pi^0} + A_{\Dn \to K^+ \pi^-} + \sqrt{2}A_{D^+ \to K^+ \pi^0} - A_{D^+ \to \Kn \pi^+} = 0.
\label{eq:DCS_sum}
\eeqn
These relations are the isospin sum rules and both are broken at the same level of $\mathcal{O} \left((m_u - m_d)/ \Lambda_{QCD} \right) \sim 1 \%$. Amplitudes involved in eq.~(\ref{eq:CF_sum},~\ref{eq:DCS_sum}) are of the same order of Cabibbo suppression and correspond to CF and DCS decay amplitudes, respectively. These sum rules could be illustrated as shown in Fig.~\ref{fig:SumRules}. 

There is another sum based on \sut\ that is particularly interesting since here CF and DCS amplitudes are mixed together:
\beq
A_{D^+ \to \Kb \pi^+} -  A_{D_s^+ \to \Kb K^+} - \frac{A_{D^+ \to \Kb K^+}}{\lambda} + \frac{A_{D_s^+ \to \Kb \pi^+}}{\lambda} +  \frac{A_{D^+ \to \Kn \pi^+}}{\lambda^2} - \frac{A_{D_s^+ \to \Kn K^+}}{\lambda^2}= 0,
\eeq
where $\lambda = \sin{\theta_{12}} $ -- is the CKM parameter and $\theta_{12}$ is the Cabibbo angle. 

Recently, the question of validity of \sut\ rules and accuracy of their approximation has become urgent as they are widely used to explain the anomalously large $CP$ violation observed in \Dn\ decays by LHCb experiment~\cite{Aaij:2019kcg}:
\beq
\Delta A_{CP} = A_{CP}(K^+K^-) - A_{CP} (\pi^+ \pi^-) = (-15.4 \pm 2.9) \times 10^{-4},
\eeq
Many papers trying to reconcile LHCb result with the Standard Model suppose that such discrepancy could be explained by the enhanced penguin amplitudes~\cite{Golden:1989qx, Hiller:2012xm, Brod:2012ud, Franco:2012ck}. In these studies it was argued that \sut\  breaking not necessary happens at the amplitude level but rather could be explained by FSI generated phases. 

% Application of \sut\ analysis to charm decays provides one with some useful insights about decay amplitudes structure. For example unexpectedly large $CP$ violation observed in \Dn\ decays by LHCb experiment~\cite{Aaij:2019kcg}:
%\beq
%\Delta A_{CP} = A_{CP}(K^+K^-) - A_{CP} (\pi^+ \pi^-) = (-15.4 \pm 2.9) \times 10^{-4},
%\eeq
%caused many papers trying to reconcile this result with the Standard Model. Some approaches~\cite{Golden:1989qx, Hiller:2012xm, Brod:2012ud, Franco:2012ck} based on approximate \sut\ flavour symmetry showed that such discrepancy could be explained by the enhanced penguin amplitudes. In these studies it was argued that \sut\  breaking not necessary happens at the amplitude level but rather could be explained by FSI generated phases. 

Structure of charm decay amplitudes could be described in terms of Wick contractions of operators of effective Hamiltonian. In this work we are particularly interested in amplitudes involving \Kn\ in the final state. In the following we adopt results for amplitude expressions obtained in~\cite{Buccella:2019kpn},
for Cabibbo-favourite amplitudes:
\beqn 
A_{\Dn \to K^- \pi^+ } &=& \frac{1}{5}(3T - 2C - K)e^{i \delta_{1/2}} + \frac{2}{5}(T+C+ \kappa ) \nonumber \\
A_{\Dn \to \Kb \pi^0 } &=& -\frac{1}{5\sqrt{2}}(3T - 2C - K)e^{i \delta_{1/2}} + \frac{3}{5\sqrt{2}}(T+C+ \kappa ) \nonumber \\
A_{D^{+} \to \Kb \pi^0 } &=& (T + C + \kappa) \nonumber \\
A_{D_s^{+} \to \Kb K^+ } &=& -\frac{1}{5}(2T - 3C + \Delta) e^{i \delta^{\prime}_{1}} + \frac{2}{5} (T + C + \kappa),
\label{eq:su3cf}
\eeqn
and for doubly Cabibbo-suppressed amplitudes:
\beqn
A_{\Dn \to  K^+ \pi^-} &=& -\frac{1}{5}(3T - 2C + K)e^{i \delta_{1/2}} - \frac{2}{5}(T+C+ \kappa^{\prime} ) \nonumber \\
A_{\Dn \to \Kn \pi^0 } &=& \frac{1}{5\sqrt{2}}(3T - 2C + K)e^{i \delta_{1/2}} - \frac{3}{5\sqrt{2}}(T+C+ \kappa^{\prime} ) \nonumber \\
A_{D^{+} \to \Kn \pi^+ }  &=& \frac{1}{5} (2T -3C + \Delta - K^{\prime}) e^{i \delta_{1/2}} - \frac{2}{5}(T+C+\kappa^{\prime}) \nonumber \\
A_{D_s^{+} \to \Kn K^{+}} &=& -(T + C + \kappa^{\prime}).
\label{eq:su3dcs}
\eeqn
Here $T$ and $C$ correspond to ``tree''-level color-connected and color-suppressed amplitudes, $K$ and $K^{\prime}$ are parameters corresponding to non-conservation of strangeness changing currents,
$\kappa$ and $\kappa^{\prime}$ -- parameters allowing \sut-breaking in CF and DCS amplitudes. Also two phases are present $\delta_{1/2}$, $\delta_{1}^{\prime}$ corresponding to $I = \frac{1}{2}$ and $I = 1$ amplitudes respectively.

Buccella {\it et~al.}~\cite{Buccella:2019kpn} performed fit to the above mentioned amplitude parameters based on observed values of $CP$ asymmetries and branching ratios for charm hadron decays, and made predictions for the strong phase difference in the $\Dn \to K^{\pm} \pi^{\mp}$ decay channel:
\beq
\delta^{+-} = (3.14 \pm 5.69)^{\circ}.
\eeq
Furthermore, this study can be extended to obtain predictions for other strong phase differences:
\beqn
\delta^{00} & = & (-3 \pm 6)^{\circ}; \\
\delta^{0+} & = & (-76 \pm 4)^{\circ} ; \\
\delta_s^{0+} & = & (108 \pm 4)^{\circ} .
\eeqn

In this paper we propose a method that allows to measure strong phase differences and amplitude ratios for final states with \Kn-meson, hence probe the validity of approaches used to explain LHCb result and in general the sum rules. Such measurements will possibly allow to identify the source and scale of \sut\ breaking in charm hadron decays.

In our previous paper~\cite{Pakhlov:2019wkn} we suggested to use semileptonic \Kn\ decays and study their time evolution to measure the complex phase between \Kn\ and \Kb\ at the production point. The method is based on the disentanglement of the production phase from the (known) $\Kn-\Kb$ mixing phase, using the later as a reference. While the method can work properly, its experimental application is a challenge due to unobserved neutrino in kaon decay. We realized that the similar sensitivity could be achieved using the standard $\Kn\to \pp$ decays, which are easy to reconstruct. In this paper we show that time evolution of the state $a\Kna + b\Kba$ decaying into \pp\ allows to extract the complex parameters $a$ and $b$ as well. In this case we utilize the $CP$-violating phase \phipm\ as a reference for the proposed measurements. Both methods are based on the rare neutral kaon effects: rare decays (semileptonic for short lived kaon component) or rare interference (between short and long lived components).

\section{Method} \label{sec:Meth}
Time evolution of the neutral kaon system could be described by Shr\"odinger equation 
\begin{equation}
i {\partial}_t
{\Kn(t) 
\choose \Kb(t)}  = 
\left({{\mathbf{M}} - \frac{i}{2} {\mathbf {\Gamma}}}\right) 
{\Kn(t)   \choose \Kb(t)} \, ,
\label{eqn:schro}
\end{equation}
where effective Hamiltonian is a sum of absorptive and dispersive parts, $\bf M$ and $\bf \Gamma$ are $2\times2$ hermitian matrices. The Hamiltonian eigenvalues could be written as follows:
\beqn
\lambda_{S,L} \equiv m_{S,\,L}-i\frac{\Gamma_{S,\,L}}{2} = \left( M_{11} - i \frac{\Gamma_{11}}{2} \right) \pm \pqk
\left( M_{12} - i \frac{\Gamma_{12}}{2}  \right) \, , 
\label{eq:Hvals}
\eeqn 
where $m_{S,\,L}$, $\Gamma_{S,\,L}$ are masses and widths of the Hamiltonian eigenstates \Ks\ and \Kl, and parameters $p$, $q$ correspond to the flavour admixtures of eigenstates defined by
\beq
\pqk^2 = \frac{M_{12}-\frac{i}{2}\Gamma_{12}}{M_{12}^*-\frac{i}{2} \Gamma_{12}^*} \, .
\eeq
Since we consider only $\pi^+ \pi^-$ final state it is convenient to use $p/q = (1+\varepsilon)/(1-\varepsilon)$, where $\varepsilon$ describes the $CP$-even component in \Kl. Then amplitudes describing evolution of initially pure flavour eigenstates in terms of \Ks/\Kl\ could be written as

\beqn
\Kna &=& \frac{(1-\varepsilon)}{\sqrt{2}} \left[ \elas \Ksa + \elal \Kla \right], 
\label{eq:ksAmp1} \\
\Kba &=& \frac{(1+\varepsilon)}{\sqrt{2}} \left[ \elas \Ksa - \elal \Kla \right].
\label{eq:ksAmp2}
\eeqn
Using these equations one can obtain for time-dependent decay rates
\begin{multline}
\mathcal{R}(t) = \frac{1 \mp 2 \mathrm{Re}(\varepsilon)}{2} |A_{K_S \to \pi\pi}|^2 \Big[ \egs + |\etpm|^2 \egl \\ \pm 2 |\etpm| \egsum \cos{(\Delta m t - \phipm)} \Big],
\label{eq:rate}
\end{multline}
where the upper (lower) sign corresponds to initial pure \Kn\ (\Kb), $\Delta m = m_{L} - m_{S}$ is a mass difference and we introduced for the amplitude ratio parameter
\beq
\frac{\left\langle \pp \left| H \right| \Kl \right\rangle}{\left\langle\pp \left| H \right| \Ks \right\rangle} = \etpm = |\etpm| e^{i \phipm}.
\eeq

The third interference term in eq.~(\ref{eq:rate}) basically allows to distinguish the initial flavour of neutral kaon.
Parameters $\etpm$ and $\phipm$ have been measured with great precision and current world averages (assuming CPT invariance) are ~\cite{Zyla:2020zbs}: $\etpm =  (2.232 \pm 0.011) \times 10^{-3}$,
$\phipm = (43.51 \pm 0.05)^{\circ}$. In the following calculations we neglect the direct $CPV$ in kaons and assume $\varepsilon = \etpm$.

Despite the smallness of indirect CPV in kaons, itself it opens interesting possibilities for flavour physics measurements~\cite{Grossman:2011zk, Bigi:1994aw, Kagan:2020vri}. 
In the following section we consider few cases of charm hadron decays that are particularly interesting for the \sut\ probe. 

\section{Strong phase difference between CF and DCS decays}
Here we consider a set of two-body decays, where both Cabibbo-favourite and doubly Cabibbo-suppressed amplitudes contribute to the final states. The following analysis could be applied to the decays $D^+ \to \Ks \pi^+$, $D_s^+ \to \Ks K^+$ and $\Lambda_c \to p \Ks$ (the later is not related to \sut\ sum rules, but is interesting on its own). Amplitudes of the final states with \Kn\ could be expressed then 
\beqn
A_{f}  & = \langle \pi \pi  |H|  \Kb \rangle + \srd e^{i \delta} \langle \pi \pi  |H| \Kn \rangle,
\label{eq:ampA}
\\
\overline{A_{f}}  & = \langle \pi \pi |H|  \Kn \rangle + \srdbar e^{i\delta} \langle \pi \pi |H| \Kb \rangle
\label{eq:ampAbar}
\eeqn
where $\delta$, $r_{f}$  -- strong phase difference and amplitude ratio for CF and DCS amplitudes. 
In general case $r_f \neq \overline{r_f}$ and present itself direct CPV in charm decays. Current world averages on $A_{CP}$ for $D^+$ and $D_s^+$ mesons are~\cite{Amhis:2019ckw}:
\beqn
A_{CP}^{D_s \to \Ks K^+} &=& (8 \pm 26) \times 10^{-4}, \nonumber \\
A_{CP}^{D^+ \to \Ks \pi^+} &=& (-41 \pm 9) \times 10^{-4}.
\eeqn
While non-zero effect was observed for $D^+$, it was noted~\cite{Belle:2012ygt} that after subtraction \Kn-\Kb-mixing contribution asymmetry is consistent with zero. So we assume no CPV thereafter $r_f = \overline{r_f}$. 

Time-dependent decay rates could be obtained by substituting \Kn (\Kb) decay amplitudes in (\ref{eq:ampA}, \ref{eq:ampAbar}) with evolution equations (\ref{eq:ksAmp1}, \ref{eq:ksAmp2}). 
\begin{eqnarray}
\mathcal{M}_{+} \equiv |A_f|^2 = \mathcal{\overline{R}} &+& r_{f} \mathcal{R} + \srd \left(  \cos{\delta} + 2 |\etpm| \sin{\delta}  \sin{\phipm} \right) \times \Big( \egs - |\etpm|^2 \egl \Big) \\ \nonumber
&+& 2 \srd |\etpm| \Big(\sin{\delta + 2|\etpm| \cos{\delta} \sin{\phipm}} \Big) \egsum \sin{(\Delta m t - \phipm)},
\label{eq:mp}
\end{eqnarray} 
\begin{eqnarray}
\mathcal{M}_{-} \equiv |\overline{A_f}|^2 = \mathcal{R} &+& r_{f} \mathcal{\overline{R}} + \srd \left(  \cos{\delta} - 2 |\etpm| \sin{\delta}  \sin{\phipm} \right) \times \Big( \egs - |\etpm|^2 \egl \Big) \\ \nonumber
&-& 2 \srd |\etpm| \Big( \sin{\delta} - 2|\etpm| \cos{\delta} \sin{\phipm} \Big) \egsum \sin{(\Delta m t - \phipm)}.
\label{eq:mm}
\end{eqnarray} 

\begin{figure}[h]
\centering
\includegraphics[width=0.8\linewidth]{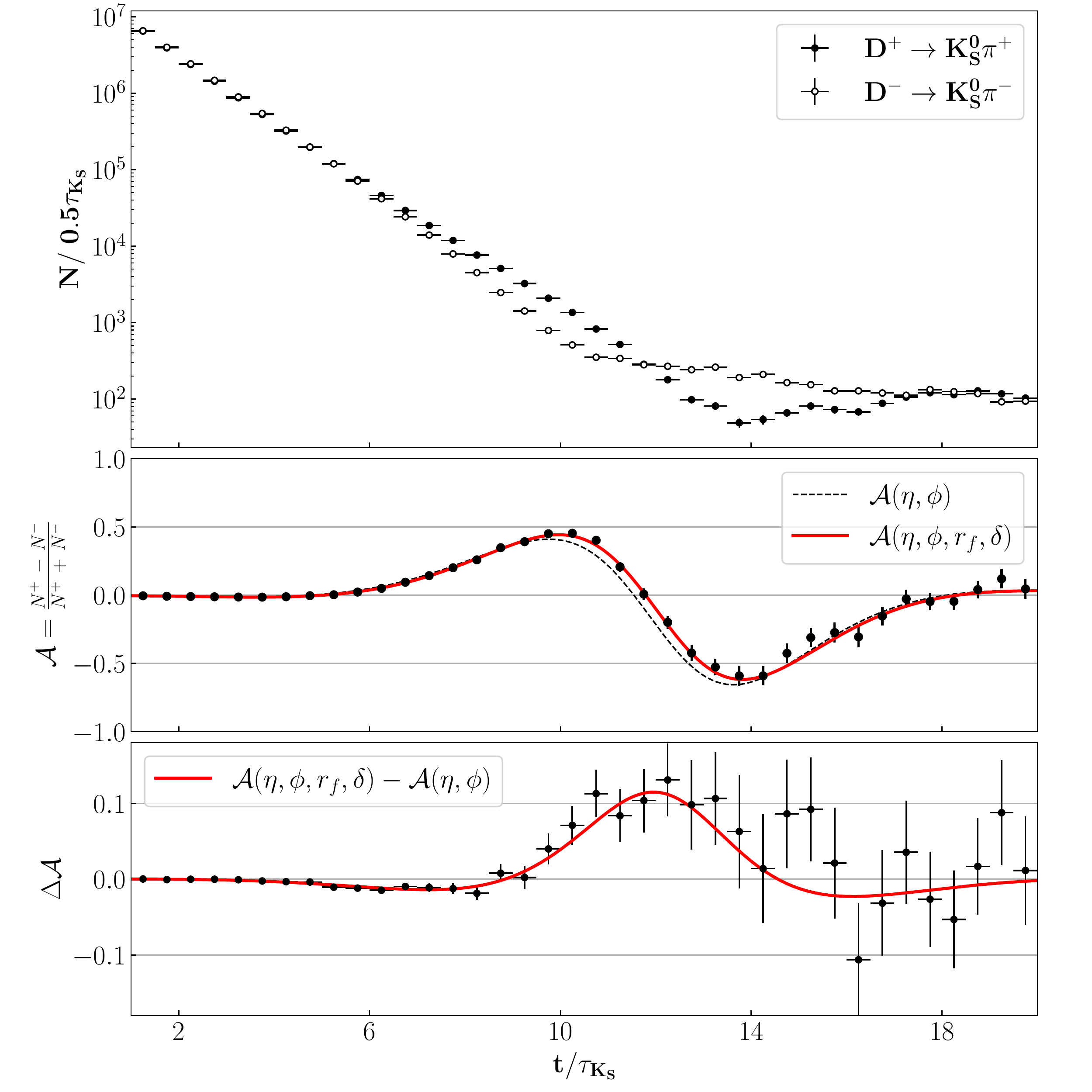}
\caption{Time-dependent decay rates for processes $D^{\pm} \to \Ks \pi^{\pm}$  {\it (top)} obtained with simulation. The bottom plots show the resulting asymmetry and the difference between asymmetries with and without DCS contribution. For this plot we used $\srd = 0.06$ and $\delta^{0+}=-76^{\circ}$.} 
\label{pic:asyExmp}
\end{figure}

These formulas demonstrate that the $\Kn\to\pp$ time-dependent decay rates depend on the initial strong phase, moreover, both sine and cosine of the strong phase enter the formula, therefore there are no trigonometrical ambiguities in this measurement. The decay rates along with asymmetry are illustrated in Fig.~\ref{pic:asyExmp}. One could see that the largest impact produced by the strong phase on resulting asymmetry falls on big lifetimes $\sim[6, 14] \, \tau_{K_S}$. 

For the \Dn-mesons situation gets complicates because of mixing. Time evolution of flavour states is described by
\beqn
\label{eq:devol}
|\Dn_{phys}(t^{\prime})\rangle &=& g_+(t^{\prime}) \Dna - \qpd g_-(t^{\prime}) \Dba, \nonumber \\
|\Db_{phys}(t^{\prime})\rangle &=& g_+(t^{\prime}) \Dba - \pqd g_-(t^{\prime}) \Dna ,
\eeqn
where $g_\pm = \frac{1}{2} \left( e^{-i \lambda_2 t^{\prime}} \pm e^{-i \lambda_1 t^{\prime}} \right)$, $\lambda_{1,2}$ -- are Hamiltonian eigenvalues defined likewise eq. (\ref{eq:Hvals}) and $t^{\prime}$ -- \Dn\ decay time. Here we assume no CPV in mixing, thus $|q/p| = 1 $. Amplitudes (\ref{eq:ampA}, \ref{eq:ampAbar}) can be rewritten as follows
\beqn
A_{f} (t, t^{\prime})  = a^{+}(t^{\prime}) \langle \pi \pi  |H|  \Kb \rangle + b^+(t^{\prime}) \langle \pi \pi  |H| \Kn \rangle,
\label{eq:ampAD0} \\
\overline{A_{f}}(t, t^{\prime})  = a^{-}(t^{\prime}) \langle \pi \pi |H|  \Kn \rangle + b^-(t^{\prime}) \langle \pi \pi |H| \Kb \rangle,
\label{eq:ampAbarD0}
\eeqn
where time-dependent coefficients are given by
\beqn
a^{+}(t^{\prime}) &\equiv& \langle \Kb \pi^0 | H |\Dn_{phys}(t^{\prime})\rangle  = 
A_{\Dn} \left[ g_+(t^{\prime}) - \srd e^{i (\delta + \phi)} g_-(t^{\prime}) \right] , \nonumber \\ %\simeq g_+(t^{\prime}) A_{\Dn}
b^{+}(t^{\prime}) &\equiv& \langle \Kn \pi^0 | H |\Dn_{phys}(t^{\prime})\rangle = A_{\Dn} \left[ \srd e^{i (\delta - \phi)} g_+(t^{\prime}) - g_-(t^{\prime}) \right],
\eeqn
and $a^{-}$, $b^{-}$ could be obtained from $a^{+}$, $b^{+}$ by substitution $\phi \to -\phi$. Combined measurements of \ddb-mixing yielded following values for mixing parameters~\cite{Zyla:2020zbs}: 
\beqn
x &\equiv& \frac{\Delta M }{\Gamma} =  (0.43^{+0.10}_{-0.11}) \%, \\ \nonumber
y &\equiv& \frac{\Delta \Gamma }{2\Gamma} = (0.60 \pm 0.06) \%, \\ \nonumber 
\phi &\equiv& Arg \qpd = (0.08 \pm 0.31) ^{\circ}.
\label{eq:mixParsD}
\eeqn

In~\cite{Kagan:2020vri} it was demonstrated that 2-dimensional distribution for the $\Dn \to \Ks \pi^0$ decay is sensitive to a set of $CP$ observables. However, it is unlikely that sufficient \Dn-lifetime resolution will be achieved in a neutral decay mode. Given the smallness of mixing parameters in \ddb-system in the next section we consider a possibility to integrate over \Dn\ lifetime and to use eq. (\ref{eq:mp},~\ref{eq:mm}) for \Dn/\Db\ as well. 

\section{Feasibility study}
In this section we estimate potential precision of proposed measurement in future experiments. It was shown in previous section that one would expect the most sensitivity for strong phase could be achieved at big lifetimes -- $[6, 14] \, \tau_{K_S}$. Based on this one could conclude that it is essential that experiment should possess large tracking detector and/or produce soft kaons. Also proper charged hadron identification is needed, since some of the final states differ by $K/\pi$ interchange.

We consider the most promising experiments the project of future Super \ctau\ factory~\cite{Bondar:2013cja} and Belle II experiment~\cite{Abe:2010gxa} that is already taking data. Both experiments possess large drift chambers ($R \sim \ 1$m) and produce relatively soft kaons, $\beta \gamma \sim 1..4$. Hadron identification in Belle II provided with TOP in barrel part and ARICH in endcaps and for the \ctau\ factory identification will be provided with FARICH detector that covers almost full solid angle. Given the spatial resolution of drift chambers $\sim 100 \mu m$, kaon life time resolution could be expected at the level of a few percent that is more than enough to perform proposed measurement. 

We perform feasibility study for the decay channels listed in Table~\ref{tab:ch}. Future Super \ctau\ factory is aiming to accumulate $10 \mathrm{ab}^{-1}$ data varying energies in c.m.s. from 3.097 GeV to 4.650 GeV. In particular $3 \mathrm{ ab}^{-1}$ will be taken at $\psi (3770)$-resonance, $1 \mathrm{ ab}^{-1}$ at $\psi(4160)$ and $1 \mathrm{ ab}^{-1}$ near $\Lambda_c^+ \Lambda_c^-$ threshold.
For the Belle II experiment main goal is $50 \mathrm{ab}^{-1}$. To estimate potential yield of charm hadrons we use $\sigma(ee \to c \bar{c}) = 1.1nb$ and fragmentation-fractions obtained in~\cite{Gladilin:2014tba, Lisovyi:2015uqa}.
There is of course ambiguity due to event selection criteria in each experiment and each particular channel. Here we used conservative estimations for number of events, assuming only $30 \%$ of event will pass the selection for Belle II experiment and $70 \%$ for \ctau\ factory, since much cleaner environment is expected there. For the \Dn\ studies we assumed $D^{*\pm}$ tagging in Belle II and semileptonic tag-side decays for $\psi(3770) \to \Dn \Db$ case at Super \ctau\ factory.  
Results are summarized in Table~\ref{tab:ch}.

\begin{figure}[ht]
\centering
\includegraphics[width=\textwidth]{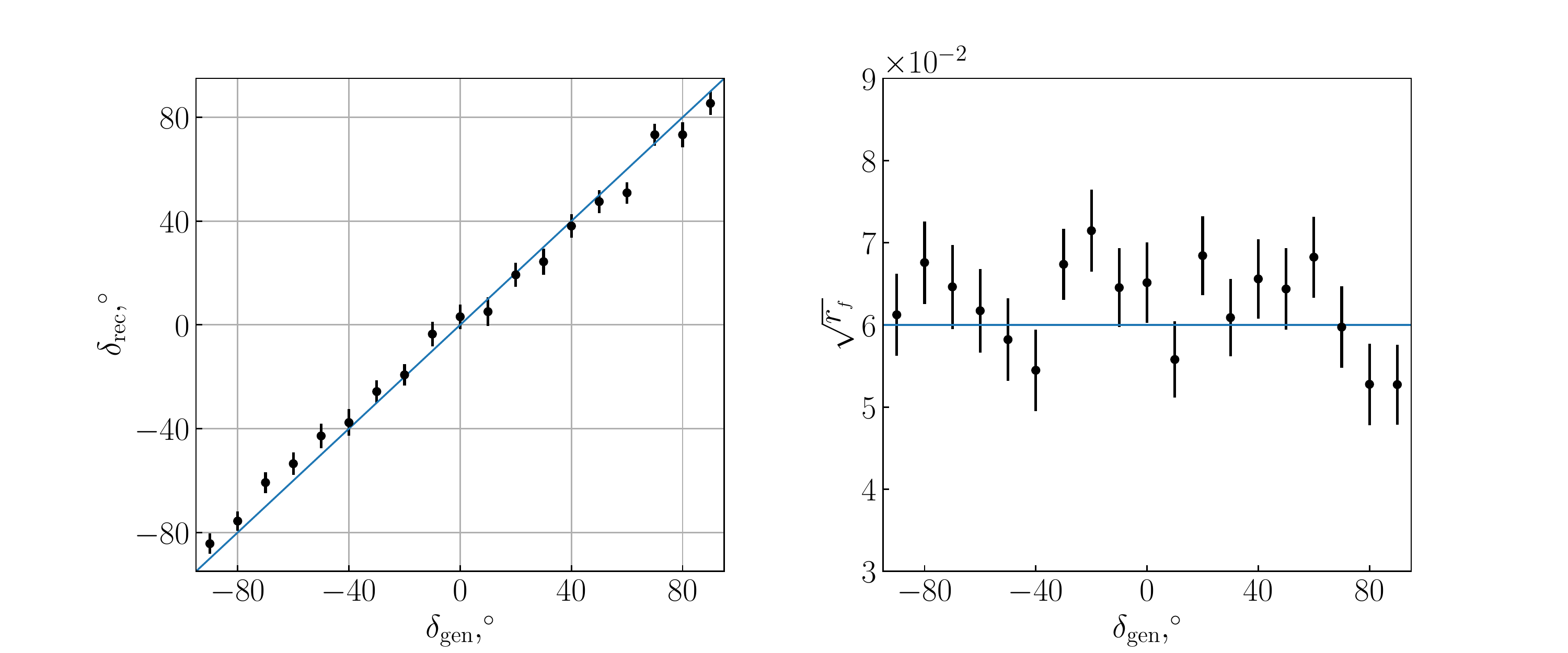}
\caption{Results of feasibility study for the decay $D^+ \to \Ks \pi^+$.  } 
\label{pic:bias50M}
\end{figure}

To confirm that there is no bias, we generate 100 MC samples of $40 \times 10^6$ events, which correspond to $D^+ \to \Ks \pi^+$ decay, each with a value of the angle $\delta$ in the $[-90^{\circ}, 90^{\circ}]$ interval with a step of $10^{\circ}$. For the two-body decay modes with \Kn\ DCS/CF amplitude ratios have not been measured yet, and could be approximated as 
\beq
r_f \equiv \left| \frac{\langle K \pi | H | D \rangle}{ \langle \overline{K} \pi| H| D \rangle} \right|^2   \simeq \left|\frac{V_{cd} V_{us}^{*}}{V_{cs}V_{ud}^{*}}\right|^2 \sim \mathcal{O}(\tan^4 \theta_{c}).
\eeq
Analogous estimation for $\Dn \to K^+ \pi^-$ slightly differs from experimental result, $r_D = (0.344 \pm 0.002) \%$~\cite{Amhis:2019ckw}. \sut-breaking terms $K$ and $K^{\prime}$ introduced in (\ref{eq:su3cf}, \ref{eq:su3dcs}) aiming to fix this small discrepancy on the amplitude level. Based on both experimental and theoretical data it is reasonable to assume other DCS/CF ratios to be of the same order $\sim \mathcal{O}(10^{-3})$.
For this test we use $\sqrt{r_f} = 0.06$ which is very close to value measured for $D^0 \to K^+ \pi^-$ decay. Each sample of MC contains time-dependent decay rates for both particle and antiparticle (see example in Fig.~\ref{pic:asyExmp}). 
For each sample we perform simultaneous unbinned maximum-likelihood fit for both time-dependent decay rates. In the fit we consider events with $t/\tau_{K_S}>1$, since kaon mixing does not contribute at low lifetimes. Fit results of one of the samples presented in Fig.~\ref{pic:bias50M}. Obtained results are in good agreement with generated values of strong phase difference and amplitude ratio.

\begin{table}
  \caption{Branching fraction and production yields}
  \begin{tabularx}{\textwidth}{m{10em} m{9em} m{10em} m{10em}}
    \hline\hline
     &  & \multicolumn{2}{c}{Estimated yield, $\times 10^6$ ($\delta$ uncertainty)}\\
    \cline{3-4}
    Channel & Branching \newline fraction, $\%$  \cite{Zyla:2020zbs} & Belle II ($50\ \mathrm{ab}^{-1}$)& Super \ctau\ \newline factory ($10\ \mathrm{ab}^{-1}$)\\
    \hline
    $D^+ \to \Ks \pi^+$ & $1.56 \pm 0.03$ & 40 ($5^{\circ}$) & 50 ($3^{\circ}$)\\
    $D_s^+ \to \Ks K^+$ & $1.46 \pm 0.04$ & 20 ($7^{\circ}$) & 40 ($5^{\circ}$)\\
    $\Lambda_c^+ \to \Ks p$ & $1.59 \pm 0.08$ & 15 ($8^{\circ}$) & 10 ($10^{\circ}$)\\
    $D^0 \to \Ks \pi^0$ & $1.23 \pm 0.02$ & 30 ($6^{\circ}$) & 20 ($7^{\circ}$) \\
    \hline\hline
  \end{tabularx}
  \label{tab:ch}
\end{table}

\begin{figure}[ht]
\centering
\includegraphics[width=\textwidth]{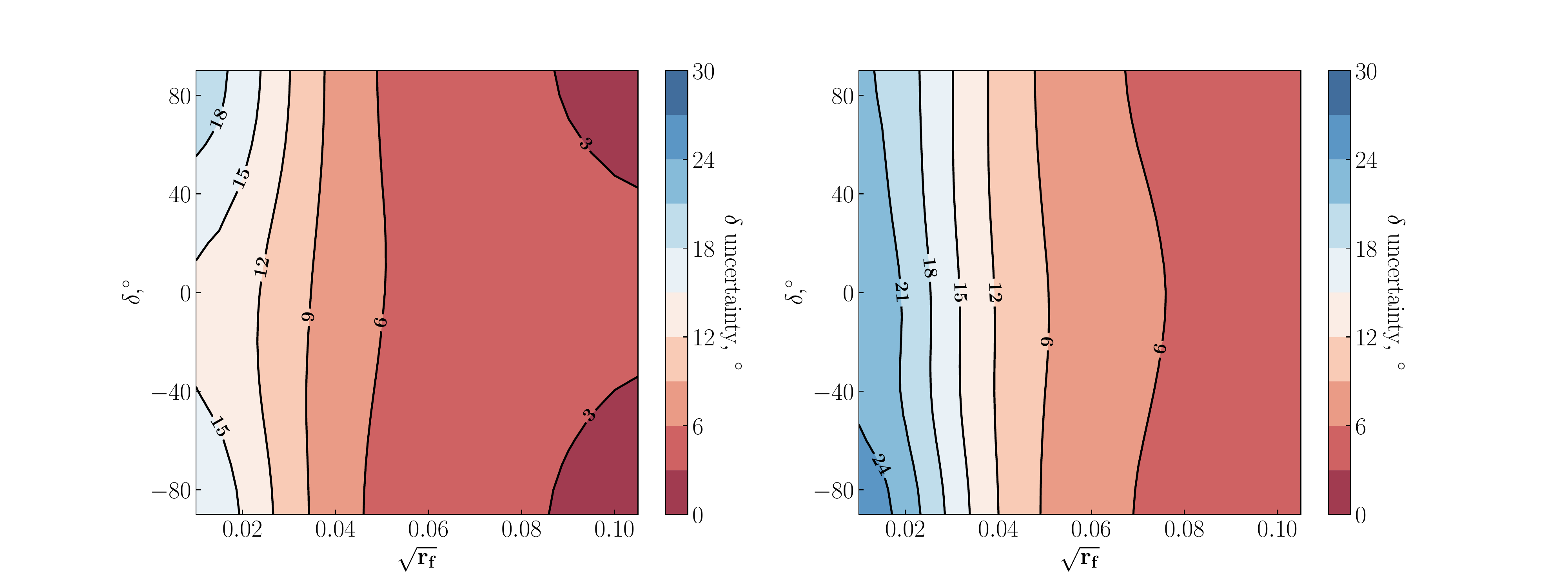}
\caption{Obtained uncertainties for the strong phase difference -- $\delta$ in $D^+$ decay {\it (left)} and in $\Dn$ decay {\it (right)}. } 
\label{pic:finalRes2}
\end{figure}

Since amplitude ratios $\sqrt{r_f}$ were not previously measured we perform a scan for its values in the interval $[0.01, 0.11]$ with step $2.5\times 10^{-3}$ simultaneously varying strong phase in the interval $[-90^{\circ}, 90^{\circ}]$. Obtained uncertainties for 20 and 50 million events, corresponding to $\Dn \to \Ks \pi^0$ and $D^+ \to \Ks \pi^+$ decays shown in Fig.~\ref{pic:finalRes2}. As one could expect we observe increasing sensitivity for strong phase with higher values of \srd . For the given amplitude ratio uncertainty in $\delta$ varies insignificantly over the range $[-90^{\circ}, 90^{\circ}]$, which is certainly the advantage of the method comparing it to usage of semileptonic \Kn\ decays~\cite{Pakhlov:2019wkn}.

Measurements for the decay $\Dn \to \Ks \pi^0$ are of great importance for the \sut\ probe. Since achieving proper \Dn\ lifetime resolution is hardly feasible we consider integration over \Dn\ lifetime -- $t^{\prime}$. For the purpose of the test MC generated distributions take into account mixing effects, but fitting $p.d.f.$ are not. 1000 pseudo experiments we performed and we found that amplitude ratio distribution turn out to be shifted at about $1\sigma$ to the higher values. Such shift in general is expected due to excess of ``wrong''-flavour kaons arised from mixing.
On the other hand strong phase measurements still proved to be in good agreement with generated values. We observed a $2^{\circ}$ bias in $\delta$, whereas the statistical uncertainty is $6^{\circ}$.

\section{Kaon regeneration}
Proposed method based on the time-dependent study of $\Kn \to \pp$ decays, however beside flavour physics parameters some other effects could contribute to decay rates. For example studies carries out in~\cite{Ko:2010mk} showed that regeneration in the environment of today's experiments could induce a bias in the $A_{CP}$ measurement up to the level of $10^{-3}$.  

In order to describe kaon propagation through matter the Hamiltonian in Schr\"odinger equation (\ref{eqn:schro}) should be modified in the following way~\cite{Good:1957zza}
\begin{equation}
    i {\partial}_t
    {\Kn(t) 
    \choose \Kb(t)}  = 
    \left({{\mathbf{M}} - \frac{i}{2} {\mathbf {\Gamma}}}\right) 
    {\Kn(t)   \choose \Kb(t)} - 
\begin{pmatrix}
\chi & 0 \\
0 & \overline{\chi}
\end{pmatrix} {\Kn(t)   \choose \Kb(t)}  ,
\end{equation}
where the second matrix describe nuclear scattering and coefficients defined as
\beq
\chi = -\frac{2 \pi N}{m}f\ \  \mathrm{and} \ \  \overline{\chi} = -\frac{2 \pi N}{m}\overline{f},
\eeq
where $f (\overline{f})$ -- are forward scattering amplitudes for $\Kn (\Kb)$, $m$ -- \Kn\ mass, $N=(\rho N_A)/M$ -- volume density of the material, $N_A$ -- Avogadro's number, $\rho$ -- mass density, $M$ -- molar mass. Strangeness conservation in strong interactions leads to inequality forward scattering amplitudes $\Delta f \equiv f - \overline{f} \neq 0$. 
The evolution of the \Ks (\Kl) beam could be expressed than
\begin{equation}
    \alpha_{S,L} = e^{-i \Sigma t} \Big[ \alpha_{S,L}^0 \cos{\left( \frac{\Delta \lambda}{2} \sqrt{1+4r^2} t \right)} \pm i \frac{\alpha_{S,L}^0 \mp 2r\alpha_{L,S}^0}{\sqrt{1+4r^2}} \sin{\left( \frac{\Delta \lambda}{2} \sqrt{1+4r^2} t \right)}
    \Big],
\label{eq:regAmp}
\end{equation}
where 
\beqn
\Sigma &=& \frac{1}{2} \left( \lambda_S + \lambda_L + \chi + \overline{\chi} \right), \nonumber \\
\Delta \lambda &=& \lambda_S - \lambda_L, \nonumber \\
\Delta \chi &=& \chi - \overline{\chi}, \nonumber \\
r &=& \frac{1}{2} \frac{\Delta \chi}{\Delta \lambda}.
\eeqn

Regeneration parameter -- $r$ is typically of the order of $10^{-2}$, so in the following calculations we use the expansion for $\alpha_{S,L}$ to the lowest order of $r$ (details could be found in Ref. \cite{Fetscher:1996fa}). It is conventional to introduce the geometrical regeneration parameter:
\beq
\zeta = r \left( 1 - e^{i \Delta \lambda \frac{Lm}{p}}\right), 
\eeq
where $p$ -- is kaon momentum and $L$ -- regenerator thickness. Amplitudes (\ref{eq:regAmp}) could be expressed than in the form: 
\beqn
\alpha_S(t) &=& e^{\frac{1}{2} (\chi + \overline{\chi}) t} e^{-i \lambda_S t} (\alpha_S^0 + \zeta \alpha_L^0 e^{-i \Delta \lambda t}), \nonumber \\
\alpha_L(t) &=& e^{\frac{1}{2} (\chi + \overline{\chi}) t} e^{-i \lambda_L t} (\alpha_L^0 + \zeta \alpha_S^0).
\label{eq:SimpRegAmp}
\eeqn
Applying equations (\ref{eq:SimpRegAmp}) recursively for each passage through matter one could account for kaon regeneration.

While accurate estimation of bias induced by regeneration should be performed for each particular experiment, here we present an estimation based on typical configurations. Since this study is mostly concerned with big kaon lifetimes, we assume that neutral kaon have to pass through a beryllium beam pipe $(\sim 1 \mathrm{mm})$ and a number of silicon layers of vertex detector. As a reference Belle II configuration was used, where silicon vertex detector consists of 6 layers ($L_{1,2} \simeq 50 \mathrm{\mu m}$ and $L_{3-6} \simeq 300\mathrm{\mu m} $).

For this test we considered only leading regeneration contribution to CF decay modes, since DCS/CF interference term is $\mathcal{O}(10^{-2})$ suppressed and DCS term $\mathcal{O}(10^{-3})$ suppressed. We used the cross sections and differences of forward scattering amplitudes obtained in \cite{Eberhard:1993nb}
\footnote{Regeneration studies in CPLEAR experiment \cite{CPLEAR:1997fwa} showed good agreement between optical model predictions and experimental results}.
Using MC simulation we found that for $1\mathrm{GeV/c}$ kaons bias in the strong phase measurement is under $4^{\circ}$. Obtained value is comparable with potential statistical uncertainty, however regeneration could be the main source of systematic uncertainty and for each particular environment studies are required. 

\section{Summary}
In this paper we presented a method to measure strong phase differences in charm hadron decay with \Kn-meson in the final state. It was shown that CPV in \Kn-\Kb\ mixing allows us to disentangle initial combination $a \Kna + b \Kba$ that arises in the presence of CF and DCS decays. In order to perform such measurement the experiment should satisfy following requirements: large tracking detector that allows reconstruct \Ks\ decays even after $10\tau_{K_S}$, sufficient statistics of charm decays -- $\mathcal{O}(10^{6}..10^{7})$ and proper charged hadron identification. The Belle II experiment and future Super \ctau\ factory are good candidates for such measurement. While LHC experiments have huge data samples of charmed mesons, some of the features of the detectors significantly reduce the possibilities of such measurements: LHCb has too short tracker, while CMS and Atlas have no particle identification.

To estimate potential precision of the method feasibility study was performed. Expected number of events was calculated for each particular channel for both experiments. Proposed measurements proved to be unbiased and free of trigonometrical ambiguity. Obtained results for statistical uncertainty (assuming $r_f \sim \mathcal{O}(10^{-3})$) are comparable with current theoretical uncertainties and uncertainties that could be obtained with semileptonic kaon decays. We also presented an estimation of regeneration contributions to proposed measurement. 

\section{Acknowledgments}

The work of P.~Pakhlov was conducted within the framework of the Basic Research Program at the National Research University Higher School of Economics (HSE). The reported study of V.~Popov was funded by RFBR, project number 19-32-90104.

\end{document}